\documentclass[aps,prl,nofootinbib,preprintnumbers,amsmath,amssymb,latexsym,array,enumerate,letter,twocolumn,superscriptaddress]{revtex4}
\usepackage{amssymb}
\usepackage{amsmath}
\usepackage{epsfig}
\usepackage{hyperref}
\usepackage{breakurl}
\usepackage{xcolor}
\usepackage{ gensymb }

\makeatletter
\def\simgt{\mathrel{\lower2.5pt\vbox{\lineskip=0pt\baselineskip=0pt
           \hbox{$>$}\hbox{$\sim$}}}}
\def\simlt{\mathrel{\lower2.5pt\vbox{\lineskip=0pt\baselineskip=0pt
           \hbox{$<$}\hbox{$\sim$}}}}
\makeatother

\def\be{\begin{equation}}
\def\ee{\end{equation}}
\newcommand{\bea}{\begin{eqnarray}}
\newcommand{\eea}{\end{eqnarray}}

\begin{document}

\title{Wobbly Dark Matter Signals at Cherenkov Telescopes \\ from Long Lived Mediator Decays}

\author{Stefania Gori}
\author{Stefano Profumo}
\author{Bibhushan Shakya}

\affiliation{Santa Cruz Institute for Particle Physics, University of California, Santa Cruz, CA 95064, USA}
\affiliation{Department of Physics, 1156 High St., University of California Santa Cruz, Santa Cruz, CA 95064, USA}

\begin{abstract}

Imaging Atmospheric Cherenkov Telescope (IACT) searches for dark matter often perform observations in ``wobble mode'', i.e. collecting data from the signal region
and from a 
corresponding background control region at the same time, enabling efficient simultaneous determination and subtraction of background. This observation strategy is possibly compromised in scenarios where dark matter annihilates to long-lived mediators that can traverse astrophysical distances before decaying to produce the gamma rays observed by the IACTs. In this paper, we show that this challenge comes with several interesting features and opportunities: in addition to signal contamination in the background control region, the gamma-ray spectrum changes with the observing direction angle and typically exhibits a hard excess at high energies. This affects signal reconstruction via subtraction of the background control region measurements in non-trivial ways. Such features represent a significant departure from the canonical picture, and offer novel handles to identify a dark matter signal and to extract underlying dark matter parameters. 
  
\end{abstract}

\maketitle

\section{Introduction and Motivation}
\label{sec:motivation}

Indirect dark matter searches with ground based Imaging Atmospheric Cherenkov Telescopes (IACTs) such as MAGIC \cite{Ahnen:2017pqx}, HESS \cite{Rinchiuso:2017pcx}, and VERITAS \cite{Zitzer:2017xlo}, which look for TeV-scale gamma rays from dark matter annihilation or decay in dark matter dense astrophysical systems, have been an integral component of the effort to unravel the nature of dark matter. The upcoming Cherenkov Telescope Array (CTA), which offers an order of magnitude improvement in sensitivity over existing telescopes, opens up further possibilities on this frontier \cite{Acharya:2017ttl,Morselli:2017ojl,Morselli:2018vaf}. 

IACTs often perform observations in the so called ``wobble mode" \cite{Fomin:1994aj}, where the signal region is observed within a solid cone at an offset angle (referred to as the ON region) from the center of the telescope's field of view, while measurements from a diametrically opposite region (with respect to the center of the field of view) of the same size and offset provides a background control region (referred to as the OFF region). In this latter region a minimal number of signal events is expected. Wobble mode observations take data from both ON and OFF regions simultaneously, providing a concurrent measurement of the background up to statistical uncertainties while minimizing systematic and time-varying effects such as atmospheric changes. A particularly promising class of targets for such observations are dwarf spheroidal galaxies (dSph), for which the dark matter distribution is sufficiently well reconstructed from its gravitational effects that the angular parameters can be appropriately optimized to facilitate signal detection \cite{Palacio:2018xim,Aleksic:2016gsu}. 

\begin{figure}[t]
\includegraphics[width=3in]{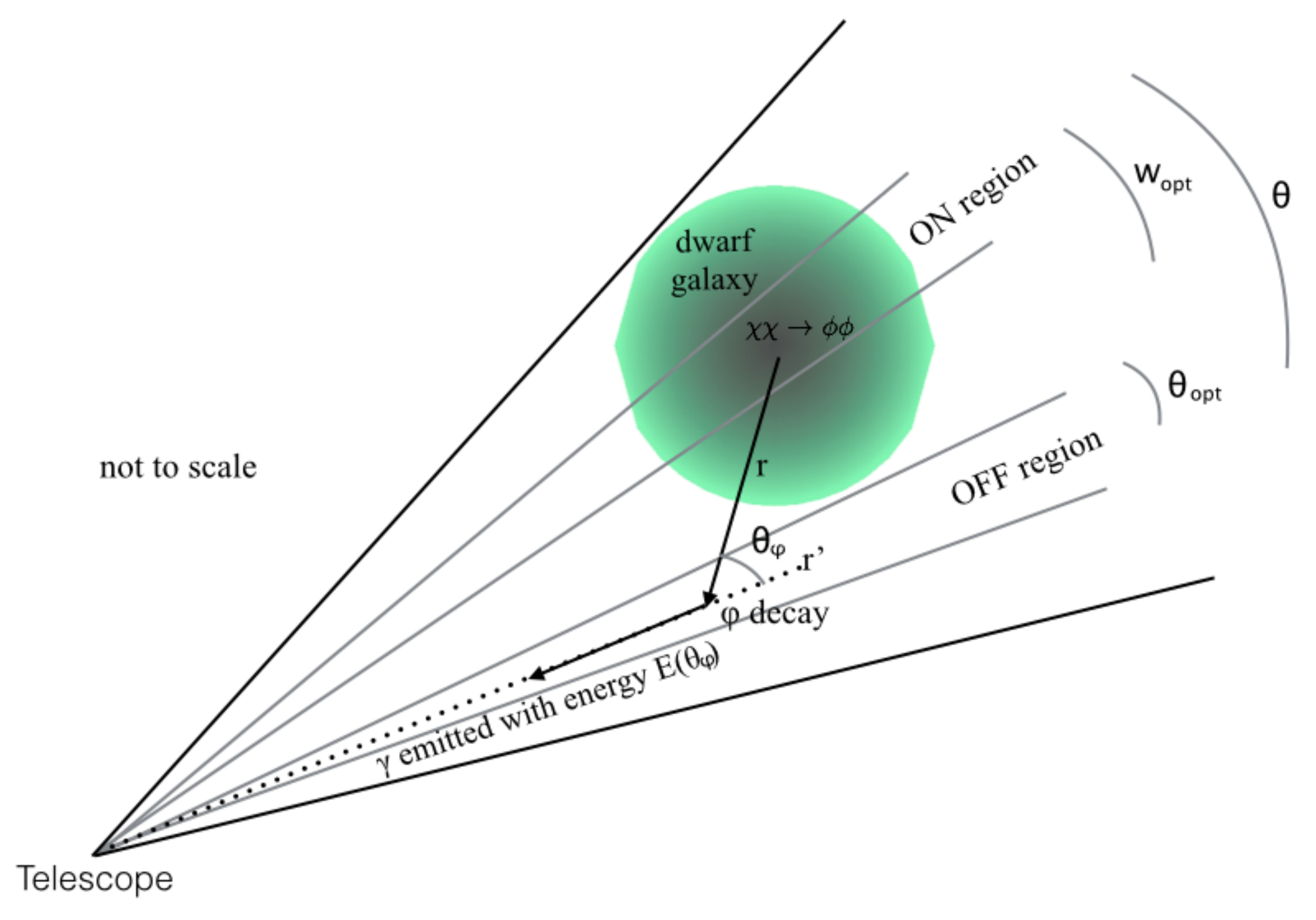}
\caption{\label{fig:cartoon} A schematic representation of the scenario studied in this paper. The telescope performs observations in  wobble mode, where the field of view contains two symmetric circular regions: the signal (ON) region, containing the signal source (here, a dwarf spheroidal galaxy), and a background control (OFF) region. We study the effects of dark matter annihilation in the ON region into long-lived mediators, $\phi$, that can propagate to and decay in the OFF region.}
\end{figure}

However, the nature of dark matter's microscopic properties remains a complete mystery, and the assumption that the distribution of visible signals of dark matter annihilation or decay follows the dark matter distribution is born out of theoretical prejudice rather than direct observational evidence. In this paper, we study how the wobble mode observation strategy is affected in scenarios where dark matter annihilates into mediator particles that are long lived and travel astrophysical distances before decaying, producing signal events in the OFF region. A schematic of this setup is shown in Fig.\,\ref{fig:cartoon}.

This study is motivated from both theoretical and experimental considerations. From a theory point of view, stringent null results from recent direct and indirect detection efforts are making it increasingly plausible that dark matter resides in some dark or hidden sector that is only very weakly connected to our visible sector (see e.g. Ref.\,\cite{Alexander:2016aln} and references therein). This opens the possibility that dark matter annihilates or decays not into Standard Model (SM) particles, but into particles that also belong to the dark or hidden sector, which then decay into the SM \cite{Pospelov:2007mp,Elor:2015bho,Profumo:2017obk,Evans:2017kti,Leane:2018kjk}. Since such ``mediator" particles are part of the hidden sector, their decay lifetime into SM states can be arbitrarily long. Here we focus on long lifetimes, corresponding to boosted decay lengths $\mathcal{O}($kpc), sufficient to produce mediator decays in the OFF regions of IACTs. Dark matter annihilation into mediators with such long lifetimes and the effect on indirect detection have been discussed earlier in the literature, for instance in Ref.\,\cite{Rothstein:2009pm,Kim:2017qaw} in the context of explaining positron excesses.  

Cosmological constraints, while in some cases extremely strict for long mediator lifetimes (see e.g. \cite{Poulin:2015opa,Poulin:2016anj,Slatyer:2016qyl,Dienes:2018yoq}), allow for such possibilities provided the abundance of such mediator particles is sufficiently small in the early Universe. This can occur in several realistic scenarios: 
Possible mechanisms include efficient mediator pair-annihilation into an additional, lighter dark-sector species, an injection of entropy after the mediator freeze-out, or exotic scenarios involving late dark matter production (e.g.\cite{DEramo:2018khz}). 

Independent of theoretical considerations, it is important to consider all plausible scenarios that can lead to departures from the standard search strategies at IACTs in order to better understand both caveats to existing results and possible new observational opportunities. 

\section{Formalism}

We consider a simplified model with the dark matter particle, $\chi$, annihilating via the process $\chi\chi\to\phi\phi$, where $\phi$ is the mediator particle. We consider $\phi$ to have a boosted decay length $l_d$ in the Galactic frame, with decay into the final state $\phi\to\gamma\gamma$\,\footnote{ We choose the final state $\gamma\gamma$ for simplicity. The analysis in this paper holds for any other decay channel with a monochromatic photon, such as a $\gamma\nu$ final state, as occurs for decays of sterile neutrinos, which can naturally be long-lived and originate from secluded sectors \cite{Cherry:2014xra,Cherry:2016jol,Shakya:2018qzg,Roland:2014vba,Batell:2017cmf}.}. The three parameters $m_\chi,\,m_\phi,$ and $l_d$ represent the free parameters of our simplified model.

In the above process, the photon spectrum is monochromatic in the $\phi$ frame with $E_\gamma=m_\phi/2$, and gets boosted into a box-shaped spectrum in the Galactic frame,
\be
\frac{dN_\gamma}{d E_\gamma}=\frac{2}{\beta_\phi m_\chi}~~~~\text{for}~~|E_\gamma-\frac{m_\chi}{2}|\leq \frac{\beta_\phi\,m_\chi}{2},
\ee
where $\beta_\phi=(1-m_\phi^2/m_\chi^2)^{1/2}$ is the mediator velocity. In the Galactic frame, there exists a correlation between the photon energy, $E_\gamma$, and the angle $\theta_\phi$ between the directions of propagation of the mediator and of the photon:
\be
E_{\gamma}=\frac{m_\phi^2/(2 m_\chi)}{1-\beta_\phi\,\text{cos}\,\theta_\phi}\,.
\label{energyangle}
\ee

For our dark matter source, we consider a dwarf spheroidal galaxy with dark matter density profile \cite{Zhao:1995cp} (see also \cite{Geringer-Sameth:2014yza,Bonnivard:2015xpq,Bonnivard:2014kza})
\be
\rho(r)=\frac{\rho_S}{\left(\frac{r+r_{\text{cutoff}}}{r_S}\right)^\gamma\left[1+\left(\frac{r}{r_S}\right)^\alpha\right]^{(\beta-\gamma)/\alpha}},
\label{dprofile}
\ee
where $r$ is the distance from the center of the dwarf galaxy, $r_S$ is the halo scale radius, inside which the dark matter density rises as $r^{-\gamma}$ down to a cutoff scale $r_{\text{cutoff}}$. This distribution reproduces the  NFW profile for $(\alpha,~\beta,~\gamma) = (1,3,1)$. The origin of the cutoff scale $r_{\rm cutoff}$ could lie in baryonic effects \cite{Fitts:2016usl}, or in self-interactions in the dark sector \cite{Tulin:2017ara}. Here, we consider an observationally motivated range for $r_{\rm cutoff},$ from $0.05$ to $0.5$ kpc \cite{Simon:2004sr}. We find that the details of the dark matter density profile are of minor importance to the central results of our study as the long lived mediators tend to smear out the signal distribution.

The quantity of interest for our study is the gamma ray flux along a line of sight at an angle $\theta$ from the direction to the dwarf galaxy as viewed by the telescope (see Fig. \ref{fig:cartoon}). To calculate this, we follow the formalism of Ref.\,\cite{Chu:2017vao}. The aforementioned flux is
\be
\Phi_\gamma(\theta,E_\gamma)=\frac{1}{4\pi}\frac{\langle\sigma v\rangle_{\chi\chi\to\phi\phi}}{m_\chi^2}\frac{dN_\gamma}{dE_\gamma}\,J_{\text{eff}}(\theta,E_\gamma).
\label{flux}
\ee
$J_{\text{eff}}(\theta)$ represents the effective source term in the direction $\theta$. For prompt $\phi$ decays, this would simply be the familiar integral over $\rho(r)^2$ along the line of sight. Due to displaced decays of the mediators, gamma rays at a particular point can originate from dark matter annihilations elsewhere in space, and the effective J-factor is instead given by
\be
J_{\text{eff}}(\theta,E_\gamma)=\!\int_0^\infty\!\!\!dr'\!\!\int_0^{2\pi}\!\frac{d\phi}{2\pi}\int_0^\infty \! \!\!dr \frac{e^{-r/l_d}}{l_d}\rho_\chi^2(\vec{x}(r,r',\phi,\theta,E_\gamma))\,.
\label{jeff}
\ee
Here, $dr^\prime$ is associated with the integral along the line of sight specified by $\theta$. Gamma rays along this line of sight, and with energy $E_\gamma$, can only be produced by mediators coming in at an angle $\theta_\phi$, as specified by Eq.\,(\ref{energyangle}). The integrals over $\phi$ and $r$ integrate over production of mediators along the cone defined by $\theta_\phi (E_\gamma)$. The factor $e^{-r/l_d}/l_d$ represents the probability distribution function for a mediator to decay at distance r from its point of production. 

This gamma-ray flux can then be integrated over the ON and OFF regions of the IACT to yield the corresponding predicted signal fluxes. In the following section, we compute this for a benchmark scenario and discuss the resulting observational effects.


\section{Results}


Let us now consider a benchmark scenario for the setup described above. We take $m_\chi=1$ TeV and $m_\phi=400$ GeV. We use as a concrete example the Draco dSph and the telescope specifications anticipated for CTA. For the Draco dark matter profile we use Eq.\,(\ref{dprofile}), with $\alpha=2.01,\beta=6.34,\gamma=0.71$, log$_{10}(r_S/\text{pc})=3.57$ \cite{Geringer-Sameth:2014yza}. We will consider the overall normalization of the signal to be a free parameter, meaning that we do not fix the precise values of $\rho_S$ and $\langle\sigma v\rangle_{\chi\chi\to\phi\phi}$ (the scaling with these parameters is trivial). For observing parameters, we use the optimized values for CTA observations of Draco as reported in \cite{Palacio:2018xim}:
\be
w_{\text{opt}}=0.6^{\degree},~~~\theta_{\text{opt}}=0.3^{\degree},
\label{optimized}
\ee
where $w_{\text{opt}}$ is the optimal offset angle (wobble distance) of the center of the ON region from the pointing direction of CTA, and $\theta_{\text{opt}}$ is the optimal angular radius of the ON and OFF regions. Since Draco is at a distance of $\approx76$ kpc \cite{Geringer-Sameth:2014yza}, for the above parameters, the OFF region is $\sim 1.2$ kpc away from the center of Draco at its closest approach. 

\begin{figure}[t]
\includegraphics[width=2.8in]{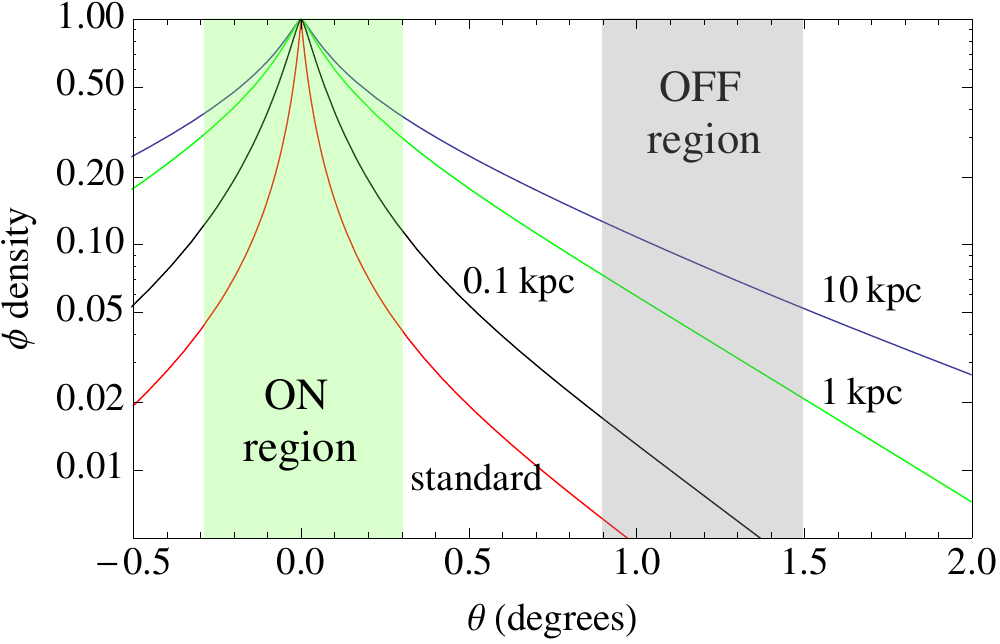}
\caption{\label{fig:rhocompare} The density of mediator particles as a function of $\theta$, normalized to 1 at the center of Draco ($\theta=0$). We present the case of instantaneous decay of the mediator in red and the cases for decay lengths $l_d=0.1,1,10$ kpc in black, green, and blue, respectively. The green and grey bands represent the optimized ON and OFF regions for CTA observations of Draco \cite{Palacio:2018xim}. }
\end{figure}

In Fig.\,\ref{fig:rhocompare}, we plot the density of mediator particles, which is the effective source term for the gamma-ray signal, as a function of $\theta$ (normalized to 1 at the center of Draco, corresponding to $\theta=0^{\degree}$) for various choices of the decay length, $l_d$. Here the mediator density is calculated along a radial line perpendicular to the direction to the detector, and we use $r_{\text{cutoff}}=0.05$ kpc for this plot. For the standard dark matter scenarios corresponding to prompt decays of mediators, this distribution reverts to the standard dark matter profile $\rho(r)^2$, for which the contribution in the OFF region is negligible (red curve), justifying the optimized choice of parameters in Eq.\,(\ref{optimized}). The effect of larger decay lengths $l_d$ is to smear out the sharp peak at the center, as mediators stream out of the central region and decay further away. This effect grows stronger for larger $l_d$. In particular, if we choose a decay length significantly larger than the $\sim 1.2$ kpc distance of the OFF region from the dwarf center (e.g. $l_d=10$ kpc), so that a significant fraction of the mediators produced in the central region reach the OFF region without decaying, we see that the contribution in the OFF region can grow to $\mathcal{O}(10\%)$ of the density at the center (blue curve). Therefore, {\em long-lived mediators can significantly contaminate the background control region of IACTs with signal events}, thus affecting the efficiency of wobble mode observations.  

\begin{figure}[t]
\includegraphics[width=2.6in]{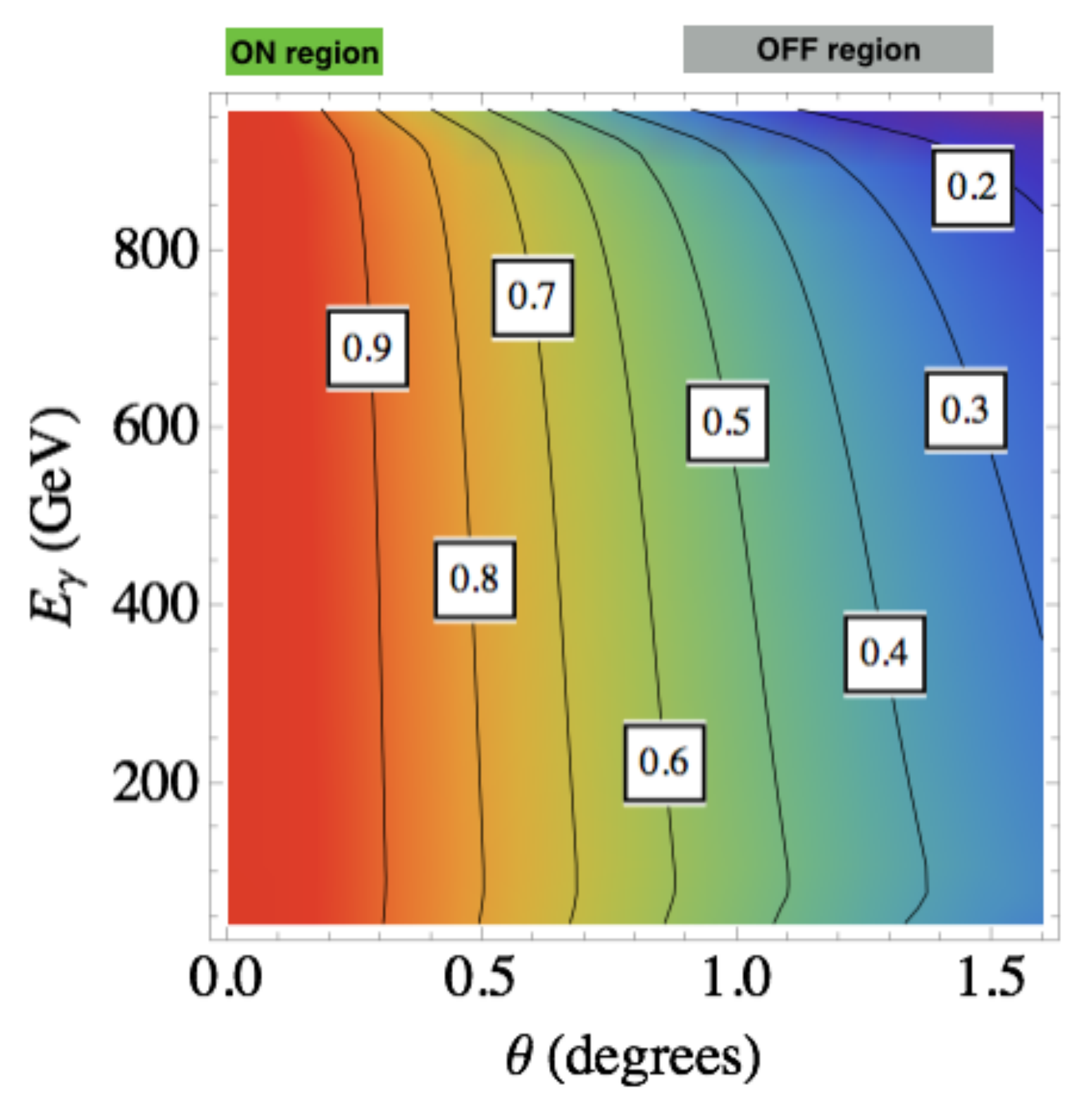}
\caption{\label{fig:photonfluxratio} Contours of the ratio $\Phi(E_{\gamma},\theta)/\Phi(E_{\gamma},\theta =0)$. i.e., flux of photons of energy $E_{\gamma}$ from direction $\theta$ relative to the center of the Draco dwarf galaxy, normalized to the flux from the center of the dwarf. }
\end{figure}

Next, in Fig.\,\ref{fig:photonfluxratio} we plot the photon flux $\Phi_\gamma(\theta,E_\gamma)$ as defined in Eq.\,(\ref{flux}) as a function of $\theta$ and $E_\gamma$, normalized to the flux from the center of the dwarf. For this plot, we have chosen $l_d=10$ kpc and $r_{\text{cutoff}}=0.5$ kpc \cite{Walker:2011zu}. This plot clearly demonstrates that the energy spectrum of the signal changes across the field of view, with the distortion getting stronger as the line of sight moves away from the center of the dwarf galaxy. There are, in particular, three distinct features: 

(i) at large $\theta$, the photon flux gets strongly suppressed at the highest energies, 

(ii) the OFF/ON ratio exhibits a peak around $E_\gamma\approx 80$ GeV, and 

(iii) the flux in the OFF region can be a sizable compared to the flux in the ON region, rising to over $50\%$ at lower energies (see the contours under the OFF region label in Fig.\,\ref{fig:photonfluxratio}). 

These effects can be understood from noting the correlation between the photon energy and the angle $\theta_\phi$ between the mediator and photon propagation directions in Eq.\,(\ref{energyangle}). The highest energy gamma rays correspond to vanishing $\theta_\phi$, i.e. to gamma rays emitted along the same direction as the mediator. For sufficiently large $\theta$ away from the galactic center, such mediators cannot originate from the central region of the dwarf (see Fig.\,\ref{fig:cartoon}), where their production rate is the greatest, resulting in a suppression of the flux at these energies. In contrast, lower energy gamma rays can be produced by mediators coming in at larger $\theta_\phi$ angles, which can originate from the core of the dwarf, enhancing the flux at these energies. The peak at $E_\gamma\approx 80$ GeV reflects the opposite effect: this energy corresponds to $\theta_\phi=\pi/2$ for the parameters chosen, so that the contribution from the center of the dwarf is received at the point on the line of sight that is also the closest point of approach (see again Fig.\,\ref{fig:cartoon}), maximizing this contribution. Due to this enhancement of intermediate/low energy photons in the OFF region, the flux in this part of the spectrum can therefore rise to over $50\%$ of the corresponding flux in the ON region.

\begin{figure}[t]
\includegraphics[width=3.0in]{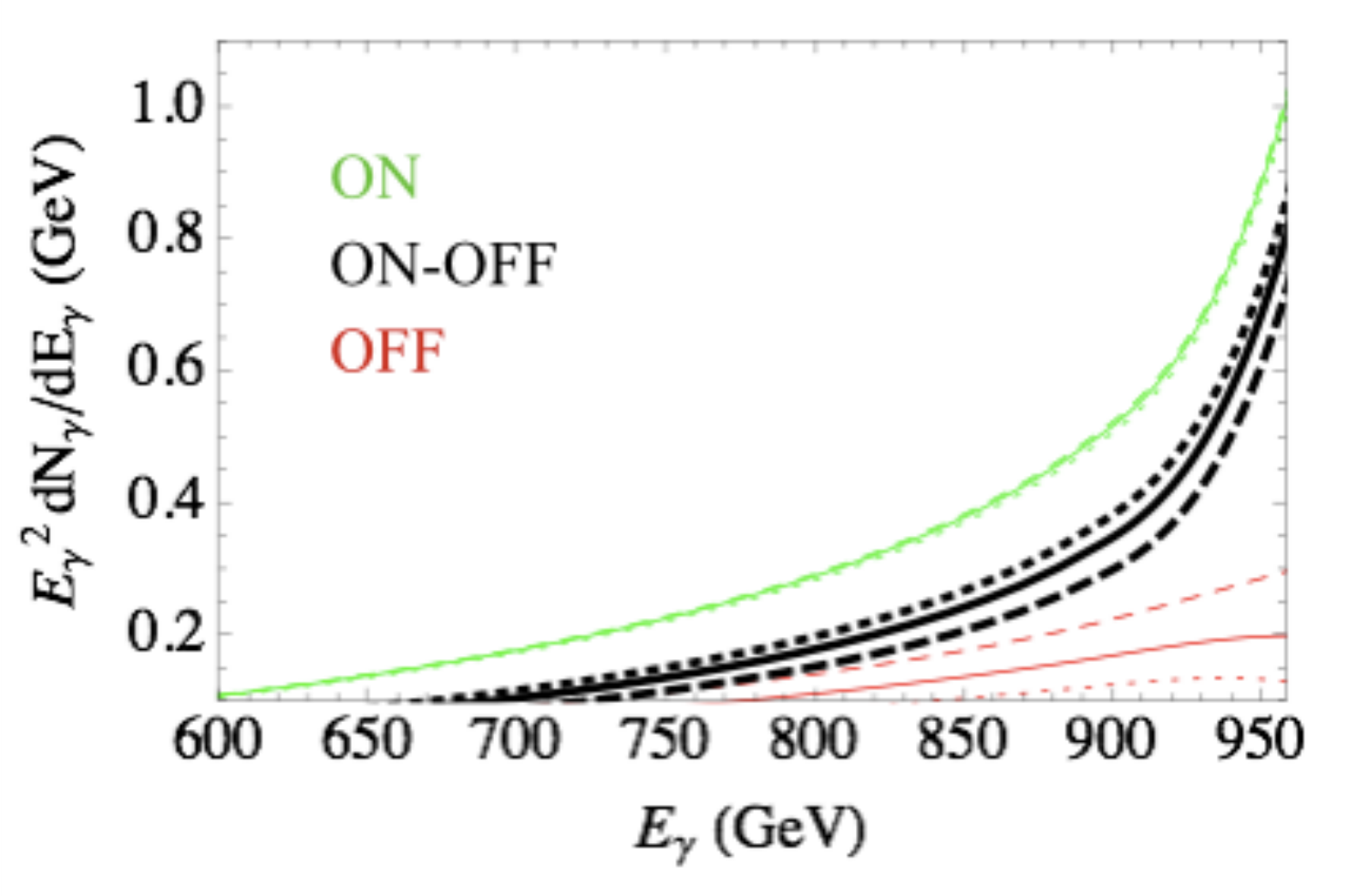}
\caption{\label{fig:onoffactual} Signal from the ON region (green), OFF region (red), and the resulting background subtracted (ON-OFF) signal (black), with an arbitrary overall normalization. The dashed and dotted curves show the corresponding curves for $w_{\text{opt}}=0.5^{\degree}, 0.7^{\degree}$ respectively.}
\end{figure}

The solid curves in Fig.\,\ref{fig:onoffactual} show the flux integrated over the ON and OFF regions described by $w_{\text{opt}},\,\theta_{\text{opt}}$ in Eq.\,(\ref{optimized}), as well as the difference between the two, choosing an arbitrary normalization. For comparison, the dashed (dotted) curves show the corresponding fluxes for $w_{\text{opt}}=0.5^{\degree} (0.7^{\degree})$.  As anticipated previously, the flux in the OFF region can be $\mathcal{O}(10)\%$ of the flux in the ON region. Note that the signal from the ON region exhibits a sharp peak at the highest energies -- these correspond to photons emitted in the direction of the mediator, so that all the photons produced from mediators emitted in the direction of the detector reach the detector. On the other hand, the spectrum from the OFF region gets suppressed at the highest energies, as these are not sensitive to mediators produced in the central region of the dwarf, as described above. Therefore, the effect of subtracting the OFF region from the ON region is that the overall signal count gets reduced, but the peak at the highest energies becomes sharper. Furthermore, the strength of this effect varies as the separation between the ON and OFF regions is varied, as indicated by the differences between the dashed, solid, and dotted curves in the figure.

Fig.\,\ref{fig:onoffactual} corresponds to $m_\chi=1$ TeV, $m_\phi=400$ GeV, and $l_d=10$ kpc. As these  parameters are varied, we find that the above plots change quantitatively, but not qualitatively. For instance, for smaller $m_\phi/m_\chi$, the mediators receive greater boost, so that the same energies correspond to smaller angles $\theta_\phi$ (as can be inferred from Eq.\,(\ref{energyangle})). The consequence of this is to stretch the curves in Fig.\,\ref{fig:onoffactual} along the $x$-axis, so that the peak gets broader. Likewise, as $l_d$ is decreased, the spectrum approaches the standard box-shaped spectrum from instantaneous decay. However, we find that the generic qualitative features discussed above -- the presence of a peak at the highest energies and distortion of the energy spectrum as a function of direction -- remain robust.

\begin{figure}[t]
\includegraphics[width=3.0in]{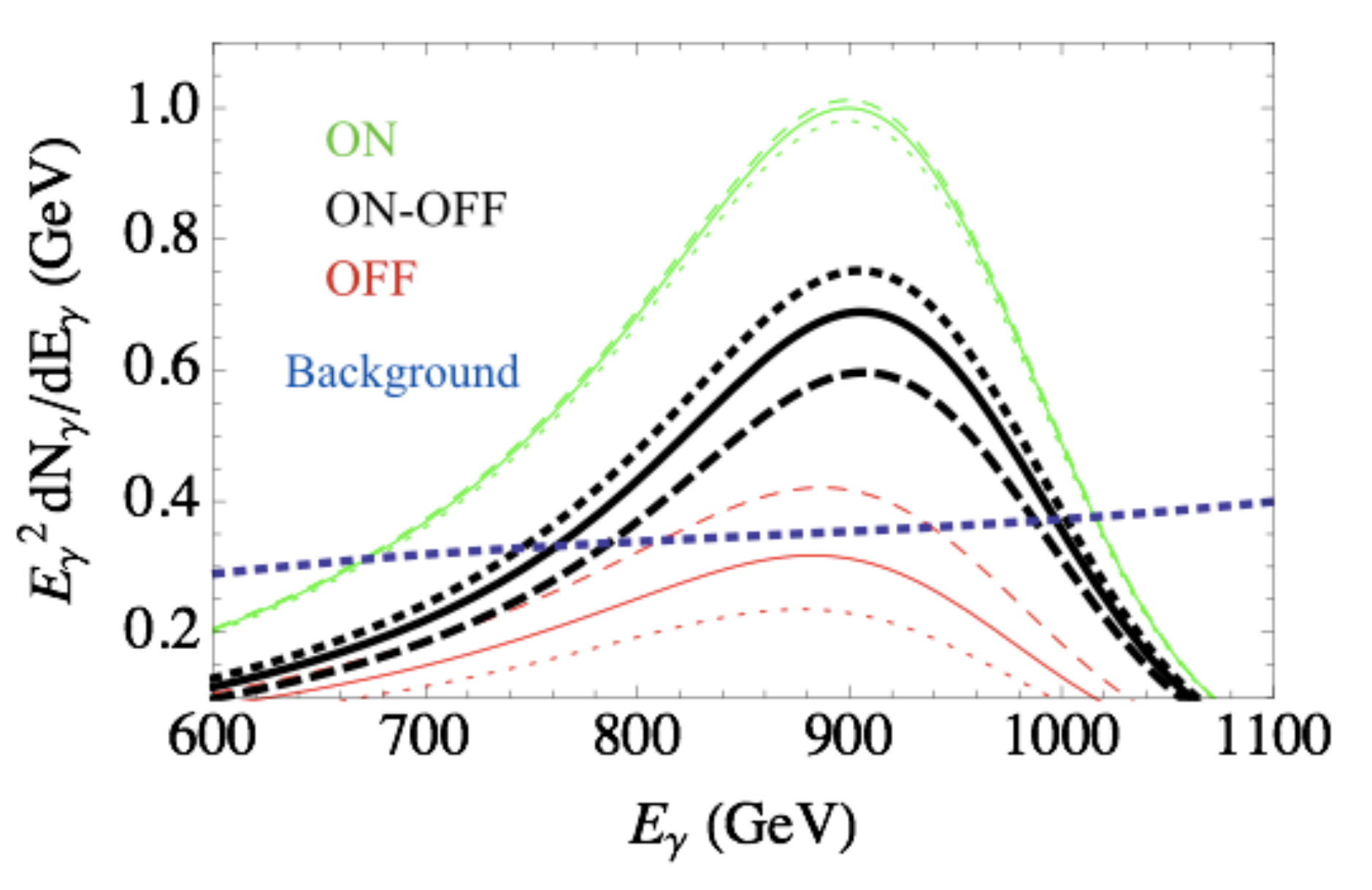}
\caption{\label{fig:onoffCTA} Same as Fig.\,\ref{fig:onoffactual}, but with CTA instrumental effects taken into account (see text for details). The most important effect is the smearing of the signal peak due to the finite energy resolution of the telescopes, which also shifts it to a lower energy. Also shown (blue dotted curve) is the expected shape of the background (arbitrary normalization).}
\end{figure}

Fig.\,\ref{fig:onoffactual} corresponds to the theoretically predicted energy spectra, without taking instrumental effects into account. We now perform a more realistic treatment of instrumental effects using CTA's publicly available data on its expected baseline performance \cite{CTA}. We use the effective collection area as a function of gamma ray energy as provided in \cite{CTA}, the relative off-axis sensitivity (accounting for reduced efficiency of the telescope towards the edges of the field of view) as shown in \cite{Palacio:2018xim}, and angular and energy resolutions of $0.05^{\degree}$ and $7\%$, respectively (approximate numbers for $E_\gamma=1$ TeV from \cite{CTA}). The resulting curves are plotted in Fig.\,\ref{fig:onoffCTA}; these are analogous to the corresponding curves from Fig.\,\ref{fig:onoffactual} (with a different arbitrary normalization) with the above instrumental effects taken into account. The most important instrumental aspect is the finite energy resolution of the telescope, which smears the sharp peak rising to the cutoff energy ($E_\gamma\approx 960$ GeV) seen in Fig.\,\ref{fig:onoffactual}. We see that the size of the peak gets reduced, the location of the peak shifts to a lower energy ($E_\gamma\approx 900$ GeV), and there are events with $E_\gamma > m_\chi$ in the spectrum. Nevertheless, the generic qualitative features from the presence of long-lived mediators -- a peak at high energies, and a direction dependent energy spectrum -- continue to persist.  

In this plot, we also show the expected shape of the background \cite{CTA}, with arbitrary normalization (dotted blue curve). Note that this background is expected to be eliminated up to statistical uncertainties when the OFF region contribution is subtracted from the ON region measurement. The size of this residual ON-OFF background depends on the size of the sample (i.e., observation time).

\section{Discussion}

In this note, we have studied how dark matter indirect detection searches at Imaging Atmospheric Cherenkov Telescopes (IACTs) observing in wobble mode are affected in scenarios where dark matter annihilates into long-lived mediators that travel large distances before decaying into visible states. The efficacy of the wobble mode hinges on a clean isolation of the signal source within the ON region, with the OFF region providing a reliable background control region. Using the parameters for the Draco dwarf spheroidal galaxy and the Cherenkov Telescope Array (CTA) as concrete, realistic examples, we studied a scenario where the mediators are sufficiently long-lived that they can travel from the ON region to the OFF region before decaying, endangering the above assumption. We demonstrated several interesting features in this setup:

\setlength{\leftmargini}{12pt}
\begin{itemize}
\item  The OFF region can receive a significant contribution of signal events ($\mathcal{O}(10)\%$ of the flux from the ON region, up to $50\%$ at low energies). This can affect the extraction of the dark matter signal from the ON region when this background control region is subtracted. 
\item A consequence of the long lifetime of the mediators is that the gamma ray spectrum features a hard excess at the highest available energies instead of being box-shaped (Fig.\,\ref{fig:onoffactual}).
\item The gamma-ray spectrum varies across the field of view of the instrument. In particular, relative to the spectrum from the center of the dwarf, the emission predicted at larger angles away from this direction feature an increasingly suppressed flux at the highest energies (Fig.\,\ref{fig:photonfluxratio}). Due to this distortion, the subtraction of the background as measured in the control (OFF) region suppresses the overall signal strength but enhances the peak feature. 
\item The above features persist even when realistic instrumental effects are taken into account. 
\end{itemize}

The features described above offer both challenges and opportunities. While the contamination of the OFF region with signal events is a cause for concern as it can significantly diminish the significance of the extracted signal, the above features also offer novel handles to identify the signal and extract the underlying dark matter parameters. The change of spectrum from a box-shape to a hard excess at high energies, which is further enhanced when the measurement from the OFF region is subtracted, can aid in identifying the dark matter contribution. The variation of the energy spectrum across the field of view of the instrument offers a challenging but tantalizing possibility -- if this effect can be reliably measured with enough statistics, it can be used to both identify the existence of long lived mediators as well as extract information about the underlying parameters such as the mediator mass and decay length. Such ``wobbly" dark matter signal features therefore offer far richer possibilities beyond canonical dark matter setups and could reveal highly non-trivial information on a possibly complex dark sector. 

Given the vastness of possibilities in dark matter and dark sectors, as well as nonstandard signatures and features beyond the canonically studied ones that our current and future instruments could be sensitive to, it becomes crucial to explore and understand the possibilities offered by such exotic scenarios in order to cast as wide a net as possible in our quest for a glimpse into the nature of dark matter.

\medskip
\textit{Acknowledgements: }We thank Jared Evans and David A. Williams for helpful conversations.  SP is partly supported by the U.S.\ Department of Energy grant number de-sc0010107. SG and BS are partially supported by the NSF CAREER grant PHY-1654502. This research has made use of the CTA instrument response functions provided by the CTA Consortium and Observatory, see http://www.cta-observatory.org/science/cta-performance/ (version prod3b-v1) for more details.

\bibliography{long_lived_mediators}

\end{document}